\documentclass[journal, twoside]{IEEEtran}
\usepackage{indentfirst}
\usepackage{graphicx}
\usepackage{amsmath}
\usepackage{amssymb}
\usepackage{amsfonts}
\usepackage{mathrsfs}
\usepackage{leftidx}
\usepackage{color}
\usepackage{amsmath}
\usepackage{arydshln}
\usepackage{amsthm}
\usepackage{ragged2e}
\usepackage{cite}
\usepackage{enumerate}
\usepackage{longtable}
\usepackage{float}
\usepackage{stfloats}
\usepackage{hyperref}
\usepackage{algpseudocode}
\usepackage{algorithm}
\usepackage[caption=false,font=footnotesize]{subfig}
\usepackage{makecell}
\usepackage{url}
\usepackage{enumitem}
\usepackage{multirow} 
\usepackage{booktabs}
\theoremstyle{plain}

\usepackage{caption}

\newcolumntype{P}[1]{>{\raggedright\arraybackslash\footnotesize}m{#1}}
\newcolumntype{A}[1]{>{\centering\arraybackslash\footnotesize}m{#1}}

\usepackage[table,usenames,dvipsnames]{xcolor}
\definecolor{aa}{RGB}{175,238,238}
\definecolor{bb}{RGB}{255,255,255}

\usepackage{bm}
\usepackage{makecell}

\begin{document}

\title{Image Steganography For Securing Intellicise Wireless Networks: ``Invisible Encryption" 

Against Eavesdroppers}

\author{Rui Meng,~\IEEEmembership{Member,~IEEE,} Song Gao, Haixiao Gao, Yinqiu Liu,~\IEEEmembership{Member,~IEEE,} Ruichen Zhang,~\IEEEmembership{Member,~IEEE,} Huishi Song, Mengying Sun,~\IEEEmembership{Member,~IEEE,} Xiaodong Xu,~\IEEEmembership{Senior Member,~IEEE,} and Ping Zhang,~\IEEEmembership{Fellow,~IEEE}

\thanks{
This work was supported in part by the National Key R\&D Program of China under Grant 2025YFF0514703; in part by the National Natural Science Foundation of China under Grant 62501066; in part by the China Postdoctoral Science Foundation under Grant 2026M793832; and in part by the S\&T Program of Hebei under Grant 262X0405D.
\textit{(Corresponding author: Huishi Song and Xiaodong Xu.)}

Rui Meng and Xiaodong Xu are with State Key Laboratory of Networking and Switching Technology, Beijing University of Posts and Telecommunications, Beijing 100876, China, and also with the Satellite Internet Testing Center, Xiong'an Aerospace Information Research Institute, Xiong'an 070001, China (e-mail: buptmengrui@bupt.edu.cn; xuxiaodong@bupt.edu.cn).

Song Gao is with State Key Laboratory of Networking and Switching Technology, Beijing University of Posts and Telecommunications, Beijing 100876, China, and also with the ZGC Institute of Ubiquitous-X Innovation and Applications, Beijing 100083, China (e-mail: wkd251292@bupt.edu.cn).

Haixiao Gao, Mengying Sun,and Ping Zhang are with the State Key Laboratory of Networking and Switching Technology, Beijing University of Posts and Telecommunications, Beijing 100876, China (e-mail: haixiao@bupt.edu.cn; smy\_bupt@bupt.edu.cn; pzhang@bupt.edu.cn).

Huishi Song is with the ZGC Institute of Ubiquitous-X Innovation and Applications, Beijing 100083, China (e-mail: songhuishi@zgc-xnet.com).

Yinqiu Liu and Ruichen Zhang are with the College of Computing and Data Science, Nanyang Technological University, Singapore (email: yinqiu001@e.ntu.edu.sg; ruichen.zhang@ntu.edu.sg).
}}

\maketitle

\begin{abstract}

As one of the most promising technologies
for \textit{intellicise (intelligent and consice)} wireless networks, Semantic Communication (SemCom) significantly improves communication efficiency by extracting, transmitting, and recovering semantic information, while reducing transmission delay.
However, an integration of communication and artificial intelligence (AI) also exposes SemCom to security and privacy threats posed by intelligent eavesdroppers. 
To address this challenge, image steganography in SemCom embeds secret semantic features within cover semantic features, allowing intelligent eavesdroppers to decode only the cover image. This technique offers a form of ``invisible encryption" for SemCom.
Motivated by these advancements, this paper conducts a comprehensive exploration of integrating image steganography into SemCom. Firstly, we assess the potential of image steganography in enhancing the security of SemCom. Secondly, we delve into traditional and DL-based image steganographic paradigms for securing SemCom. Thirdly, we further discuss six image semantic steganography strategies. Fourthly, we present a case study to illustrate the effectiveness of coverless steganography SemCom. Finally, we propose future research directions for image steganography SemCom.

\end{abstract}


\section{Introduction}
The \textit{intellicise (intelligent and concise)} wireless network is an innovative paradigm designed to address complex challenges in modern communication systems by harmonizing intelligence and simplicity \cite{zhang2026towards}. Rooted in the integration of multidisciplinary theories, including information theory, artificial intelligence (AI), complex science, and system theory, it employs advanced methodologies such as basic probability, fuzzy measurement, and logical reasoning to explore fundamental principles of information and networks.
As one of the most promising technologies for intellicise wireless networks, Semantic Communication (SemCom) is an emerging communication paradigm that focuses on the semantics of the data itself during information transmission rather than the precise transmission of bit streams. By comprehending, compressing, and conveying the inherent meaning of information, it reduces redundant data transmission, thus enhancing communication efficiency and supporting more intelligent interactions \cite{guo2024survey,meng2025survey}.

However, with the development of SemCom, a series of complex security and privacy threats that traditional communication systems have never encountered have emerged. Compared with the security challenges faced by traditional wireless communication, such as eavesdropping attacks and jamming attacks, SemCom is confronted with more severe security threats due to the deep integration of AI and communication technologies \cite{guo2024survey}, such as semantic knowledge base poisoning \cite{shen2024secure}, gradient leakage \cite{yang2024secure}, and intelligent eavesdropping \cite{guo2024survey}.


To effectively address the security and privacy threats in SemCom, researchers have explored numerous defense technologies at the levels of model training, model transfer, and semantic information transmission \cite{meng2025survey}. Data cleansing ensures the purity of model training by detecting and removing malicious samples from the semantic knowledge base \cite{shen2024secure}. During the model training process, adversarial samples are proactively introduced to enhance the model's robustness against semantic adversarial attacks \cite{yang2024secure}. Differential privacy-based noises are added to semantic data to prevent attackers from inferring individual information from the output results \cite{yang2024secure}. Cryptography converts plaintext semantic data into ciphertext semantic data to safeguard confidentiality and integrity. 
Blockchain achieves the immutability and traceability of semantic data through distributed ledgers and smart contracts \cite{meng2025survey}. Model compression simplifies the model structure through techniques such as pruning, quantization, and knowledge distillation, indirectly improving security \cite{shen2024secure}. Physical layer security leverages the physical characteristics of wireless channels to provide endogenous semantic security guarantees. For instance,  Feng \textit{et al.} \cite{feng2026toward} proposed a covert semantic satellite communication framework, where sensitive semantic information is selectively protected through adaptive power allocation to reduce the probability of detection.


In addition to the aforementioned defense methods, image steganography \cite{rustad2023digital,hu2024learning,song2024survey} offers a promising approach for securing SemCom by embedding secret images into cover images, which effectively conceals encryption behavior to against semantic eavesdropping. This technique, often referred to as ``invisible encryption," holds significant potential for enhancing the security of SemCom systems. Here, invisibility involves visual imperceptibility, semantic concealment, and resistance to steganalysis, which can be evaluated using image quality, semantic similarity, and detection performance. Invertible Neural Networks (INNs) \cite{tang2024secure,tang2025towards,li2024multi} and Generative Adversarial Networks (GANs) \cite{huo2025image} have been employed to embed secret semantic features into cover semantic features, ensuring reliable extraction and maintaining covertness throughout SemCom.
Nevertheless, research on image steganography-based SemCom is still in its nascent stages. The full potential of leveraging image steganography techniques to secure SemCom has yet to be thoroughly explored. Several key questions remain unanswered:

\begin{itemize}
    \item Q1: What makes image steganography particularly suitable for securing SemCom?
    \item Q2: How can image steganography be effectively integrated with SemCom to defend against eavesdroppers?
    \item Q3: What directions can further advance image steganography for securing SemCom?
\end{itemize}

In light of these questions, we evaluate the feasibility of applying image steganography to SemCom. The main contributions of this work are as follows:
\begin{itemize}
\item We present an overview of image steganography techniques, subsequently analyzing the potential of image steganography in securing SemCom.
\item We detail image steganography paradigms for securing SemCom, comparing Convolutional Neural Network (CNN)-based, GAN-based, and INN-based image steganography joint source-channel coding (JSCC) models, and listing image semantic steganography strategies. Moreover, we present a case study to show its effectiveness.
\item We envision future research directions for image steganography SemCom.
\end{itemize}

\section{Overview of Image Steganography for Securing SemCom}

\subsection{What is Image Steganography?}
\begin{figure}
\centering

\subfloat[Basic architecture of the image steganography system, where secret images are confidential images that need to be protected or transmitted; cover images appear to be ordinary images on the surface but possess high complexity (rich textures, diverse colors, etc.) to conceal information without causing visual anomalies; and stego images are the cover images in which secret information is embedded \cite{rustad2023digital}.]{\includegraphics[width=85mm]{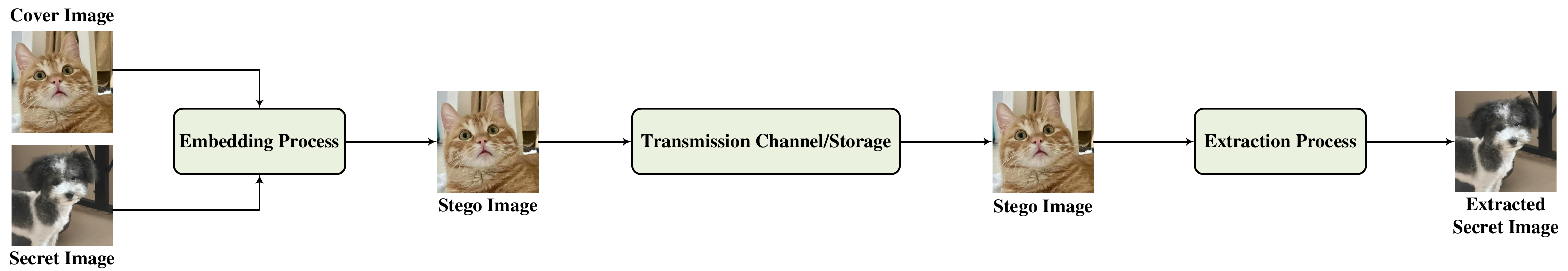}\label{fig11}}
\hfill

\subfloat[CNN-based image steganography JSCC model.]{\includegraphics[width=85mm]{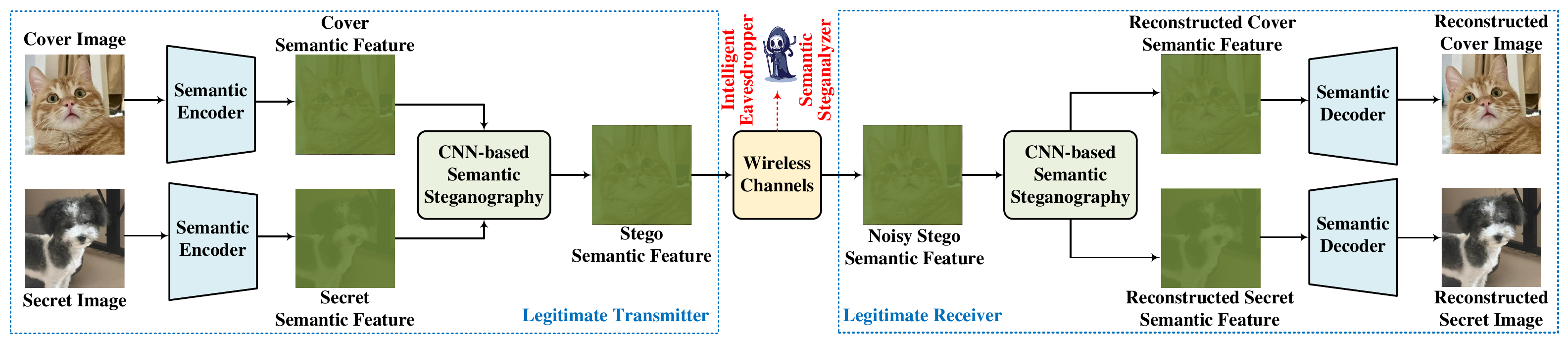}\label{fig12}}
\hfill
\subfloat[GAN-based image steganography JSCC models.]{\includegraphics[width=85mm]{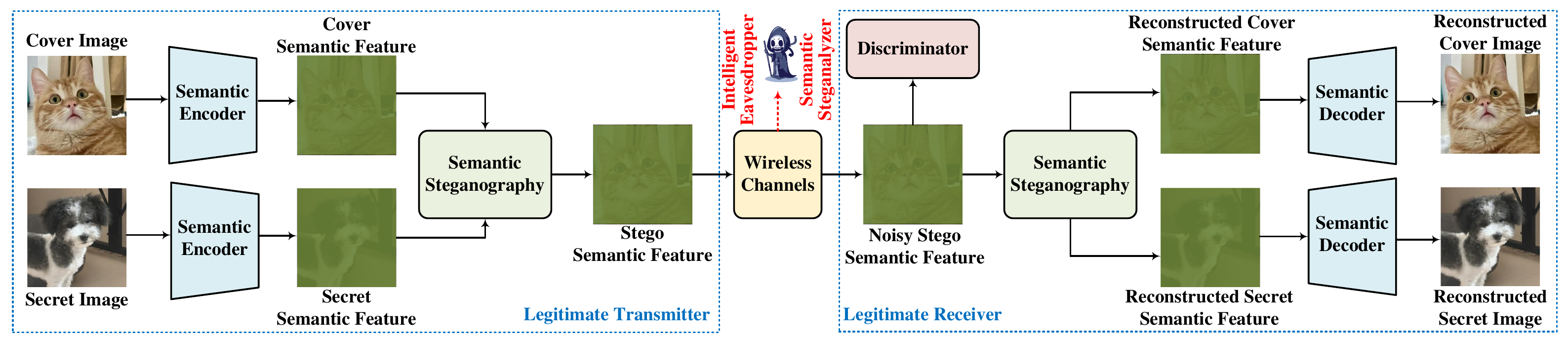}\label{fig13}}
\hfill
\subfloat[INN-based image steganography JSCC model.]{\includegraphics[width=85mm]{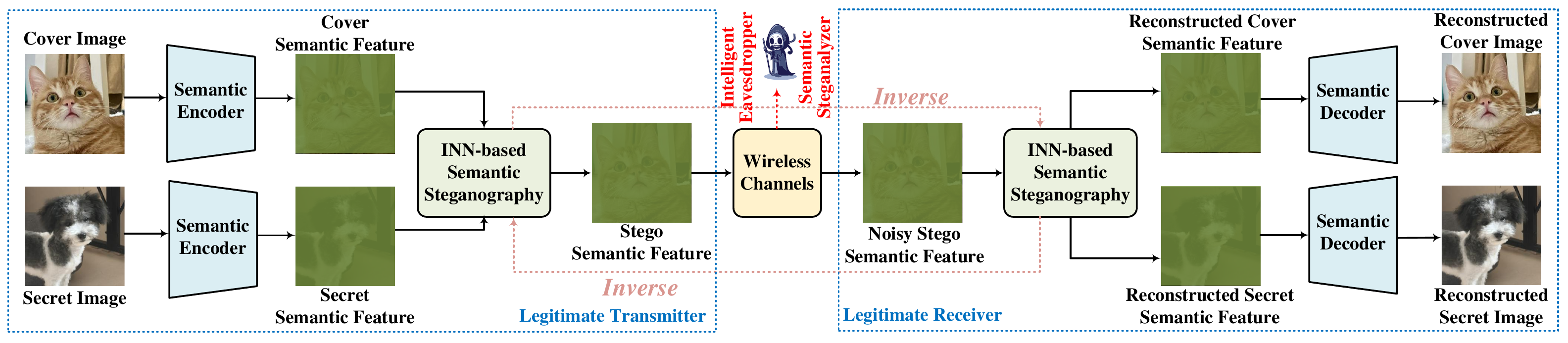}\label{fig14}}
  
\caption{Illustration of image steganography SemCom systems, where (a) presents the basic architecture of the image steganography system, (b) CNN-based image steganography JSCC model has lowest training time and model size but worst secret semantic capacity and transmission quality, (c) GAN-based image steganography JSCC model has the best anti-eavesdropping ability through adversarial training but the most parameters, and (d) INN-based image steganography JSCC model has the highest secret semantic capacity and transmission quality but longest training time.}
\label{fig1}
\end{figure}

In today's digital era, the demand for data security has driven the development of security systems: cryptography systems and information hiding systems. Cryptography systems focus on the unreadability of information, while information hiding systems emphasize concealing data to create the appearance of no secret data existing. Data hiding systems encompass watermarking and steganography techniques. Watermarking aims to prove ownership or integrity of carriers, or enable tamper-proof authentication. In contrast, steganography prioritizes hiding the existence of secret information, ensuring that third parties remain unaware of the presence of secret data \cite{hu2024learning}. The basic architecture of image steganography is illustrated in Figure \ref{fig11}. At the transmitter, a secret image is embedded into a cover image through a steganographic encoder to generate a visually inconspicuous stego image. After transmission, the legitimate receiver employs a corresponding decoder to extract the secret image.

\subsection{How to Achieve Image Steganography?}
Image steganography is realized through embedding processes, which are mainly divided into two categories:

\subsubsection{Traditional Image Steganography} 
Traditional methods can be classified into spatial-domain and transform-domain approaches according to the embedding domain. Spatial domain methods utilize manually designed hiding rules and distortion functions to perform operations in the pixel domain, enabling the embedding and extraction of secret information, where representative techniques include least significant bit substitution, pixel-value differencing and so on. Transform domain methods transform images from the pixel domain to the frequency domain and embed secret information into high-frequency regions to minimize interference with the original cover image. Transform methods involve Discrete Wavelet Transform (DWT), Discrete Cosine Transform (DCT), integer transforms, and so on \cite{rustad2023digital}.

\subsubsection{Deep Learning (DL)-Based Image Steganography} With advancements in DL-based steganalysis techniques, traditional image steganography methods rely on manually designed embedding rules, limiting their adaptability to different image contents and steganalysis techniques. To address this challenge, DL-based image steganography has emerged, which employs neural networks to automatically learn the embedding and extraction of secret information through an encoder-decoder structure, enabling adaptive feature representation and improved concealment compared with traditional methods \cite{hu2024learning}. SemCom is typically implemented via JSCC, which can be integrated with DL-based image steganography. In this way, secret information can be embedded into the semantic feature domain to achieve secure transmission. Figures \ref{fig12}, \ref{fig13}, and \ref{fig14} show CNN-, GAN-, and INN-based image steganography JSCC architectures, respectively.

\paragraph{CNN}
Compared with fully connected neural networks, CNNs use local connections and shared convolution kernels to reduce the number of parameters and extract image features such as edges and textures. These features can be used to determine suitable representations and locations for secret information embedding.
As shown in Figure \ref{fig12}, CNN-based image steganography JSCC mainly consists of semantic encoders, CNN-based semantic steganography modules, a wireless channel, and semantic decoders. The semantic encoders convert the cover and secret images into cover and secret semantic features. At the transmitter, the secret semantic features are embedded into the cover semantic features to generate stego semantic features. After wireless transmission, the receiver extracts the reconstructed cover and secret semantic features from the noisy stego semantic features. The semantic decoders then recover the cover and secret images. During training, a noise layer can be introduced to simulate channel noise, fading, and interference, improving the recovery performance under different channel conditions.

\paragraph{GAN}
GAN consists of a generator and a discriminator trained through the minimax objective $\min_G \max_D V(D,G)$. The generator produces samples that approximate the distribution of real data, while the discriminator distinguishes generated samples from real ones.
In GAN-based image steganography JSCC shown in Figure \ref{fig13}, the generator embeds secret semantic features into cover semantic features to produce stego semantic features. The discriminator distinguishes the stego semantic features from the original cover semantic features and feeds the detection result back to the generator. Through adversarial training, the generator learns to produce stego features that are difficult to detect. For instance, Huo et al. \cite{huo2025image} introduced two discriminators to improve security and semantic extraction accuracy. The first discriminator protects secret information in the semantic feature domain, while the second prevents its exposure after image reconstruction.

\paragraph{INN}
In INNs, the input to each layer can be fully recovered from its output, meaning there exists a well-defined mathematical inverse operation for every transformation.For a forward mapping $y=f(x)$, its input can be recovered through the inverse mapping $x=f^{-1}(y)$. This property allows the embedding and extraction processes to be implemented within the same invertible structure and reduces information loss during secret recovery.
In the INN-based image steganography JSCC architecture shown in Figure \ref{fig14}, the forward process combines the cover and secret semantic features to generate stego semantic features, while the inverse process recovers the secret semantic features from the received features. Tang et al. \cite{tang2024secure} proposed an INN-based signal steganography framework that uses multiple invertible blocks and additive affine transformations to embed secret signals. Its security was further evaluated under glass-box, closed-box, GAI-assisted glass-box, and GAI-assisted closed-box eavesdropping scenarios \cite{tang2025towards}. Li et al. \cite{li2024multi} proposed an INN-based multimodal hiding and deception method, which generates virtual semantic information to mislead eavesdroppers and hides the original semantic information of one modality in another modality. Legitimate users can recover the hidden information through the inverse process.


\subsection{What Can SemCom Benefit from Image Steganography?}
Compared with existing SemCom encryption techniques, image steganography offers the following advantages:
\begin{itemize}
\item \textbf{Concealing Encryption Traces:} Data processed by traditional encryption methods becomes unreadable, directly revealing the act of semantic encryption. In contrast, steganography can integrate secret information into the semantic feature space, enabling the stego semantic features to convey both normal semantic information and additional secret semantic information, thus achieving the concealment of encryption \cite{huo2025image}.
\item \textbf{End-to-End Implementation:} Traditional encryption requires separate encryption and decryption modules, which not only increase system complexity but also introduce additional latency. In contrast, steganography can be deeply integrated with the encoding and decoding architecture of SemCom, allowing the receiver to extract secret semantic information without the need for additional processing steps, thereby improving communication efficiency \cite{tang2024secure,tang2025towards}.
\item \textbf{Defense Against Intelligent Eavesdroppers:} Traditional encryption methods generate ciphertext that, even if it cannot be directly decrypted, allows intelligent eavesdroppers to infer sensitive data through model inversion attacks (MIA) and Generative AI (GAI) techniques. In contrast, steganography embeds secret semantic features into cover semantic features without altering the cover image's visual semantics. This makes it difficult for intelligent eavesdroppers to detect the presence of hidden information, let alone extract or infer sensitive content, thereby defending against them \cite{tang2025towards}.
\item \textbf{High Flexibility and Compatibility:} Steganography exhibits excellent flexibility and can adaptively adjust to the requirements of different scenarios. In bandwidth-constrained situations, steganography can adjust the steganographic capacity to prioritize the quality of semantic transmission. When facing severe security threats, more robust steganography methods can be employed to enhance wireless security. Additionally, steganography can complement traditional encryption technologies, forming a ``dual protection" mechanism. Even if traditional encryption methods are compromised, attackers still need to perform complex steganalysis to access the secret content, providing more reliable security for SemCom.
\end{itemize}

\begin{figure*}
\centering
\includegraphics[width=1\textwidth]{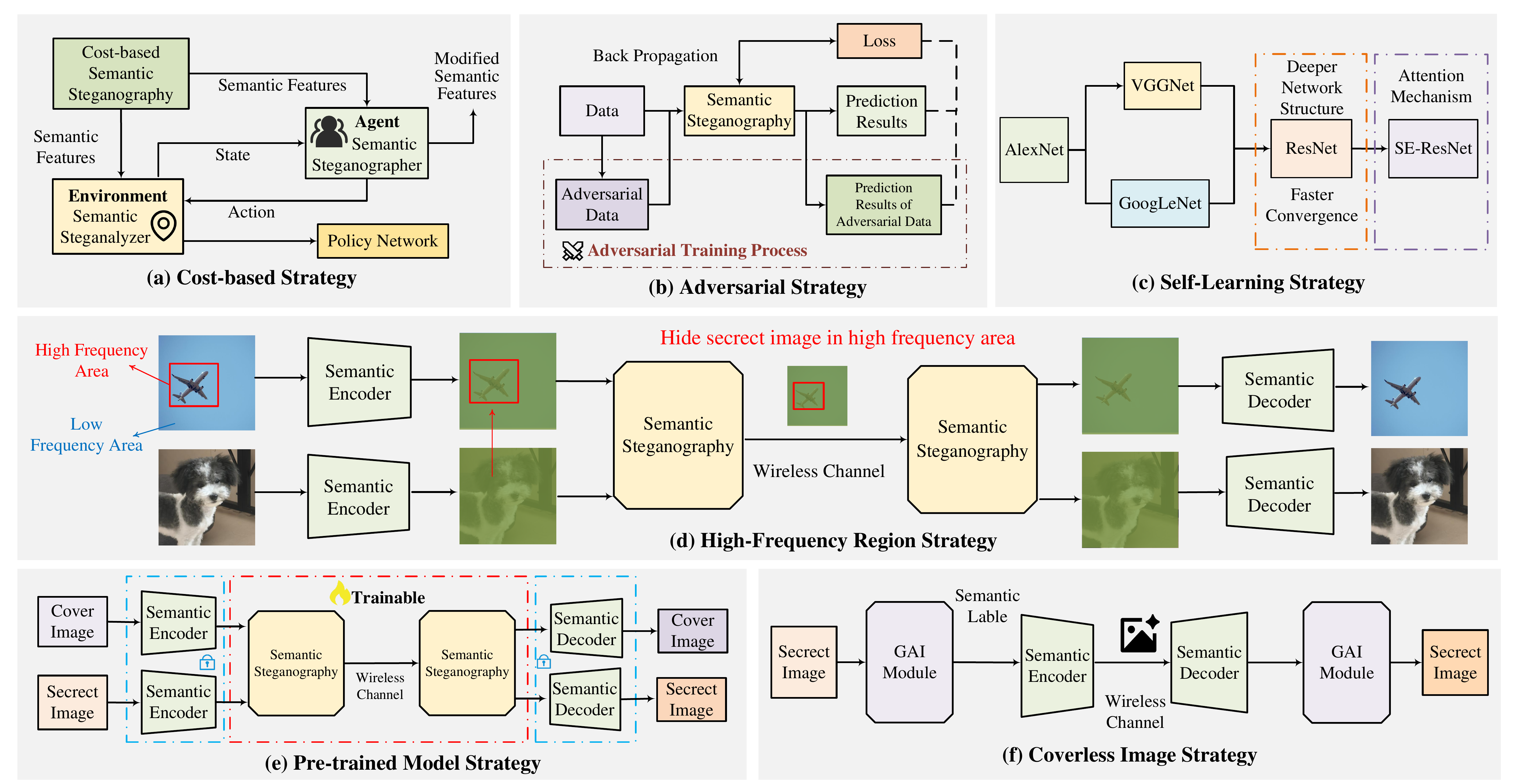}
\caption{Illustration of six image semantic steganography strategies.
(a) Cost-based strategy leverages reinforcement learning to optimize embedding decisions with minimal cost, making it ideal for scenarios with limited steganographic capacity.
(b) Adversarial strategy integrates adversarial examples during training to enhance resilience against detection by semantic steganalyzers.
(c) Self-learning strategy utilizes end-to-end DL models to enable autonomous semantic steganography, eliminating the need for manual preprocessing or post-processing steps.
(d) High-frequency region strategy embeds secret semantic information into high-frequency domains, leveraging regions with rich cover semantic features to evade semantic steganalysis.
(e) Pre-trained model strategy avoids time-consuming model retraining, thus enhancing steganographic efficiency.
(f) Coverless strategy generates stego images directly through GAI models, bypassing the requirement for cover images.
}
\label{figure2}
\end{figure*}

\section{Image Steganography Strategies For Securing SemCom}
Image steganography in SemCom needs to conceal secret information from eavesdroppers while maintaining reliable recovery over wireless channels.
As illustrated in Figure \ref{figure2}, we present six strategies to support the training of image steganography JSCC models, which are discussed in detail as follows.

\textbf{Cost-based Semantic Steganography:}
Cost-based semantic steganography can be designed using a semantic distortion minimization strategy to modify stego semantic features with least detectable artifacts. 
By introducing a reinforcement learning framework, the optimal trade-off between semantic steganographic security and semantic distortion can be achieved through dynamic game interactions and intelligent decision-making \cite{tang2021automatic}. 
Specifically, the agent acts as a semantic steganographer, taking the semantic distortion levels as actions, aiming to learn the optimal steganography strategy associated with semantic steganography costs. 
Meanwhile, the environment simulates a semantic steganalyzer, providing feedback to the agent’s semantic embedding actions: negative rewards for modifications easily detectable and positive rewards for imperceptible ones. 

\textbf{Adversarial Semantic Steganography:}
Adversarial examples are generated by adding imperceptible perturbations to original images, which can cause neural network-based classifiers to produce erroneous judgments. Inspired by this, adversarial semantic steganography embeds secret semantic features into cover semantic features while leveraging adversarial perturbations to enhance the security of stego semantic features, enabling them to resist detection by semantic steganalyzers. Specifically, we can leverage the generator in GANs to learn universal adversarial perturbation patterns, enabling it to automatically generate perturbations tailored to diverse semantic features, thereby enhancing the deceptive nature of stego semantic features.


\textbf{Self Learning-based Semantic Steganography:}
The self learning-based semantic steganography strategy leverages the end-to-end optimization capability of deep learning models to automatically learn the optimal way of embedding secret semantic features into cover semantic features without relying on manually predefined rules. Some typical models include AlexNet, VGGNet, GoogleNet, ResNet, etc.
In addition, the attention mechanism can dynamically learn importance weights of different channels, thereby optimizing the deep semantic representation of images and improving the quality of stego semantic features as well as the accuracy of secret semantic feature extraction.

\textbf{High-frequency Region Semantic Steganography:}
The low-frequency components of images are associated with smooth regions where the human visual system is particularly sensitive. Consequently, even minor alterations in these areas are easily noticeable by steganalyzers. In contrast, high-frequency components correspond to rapidly changing areas, such as textures and edges, which are often rich in semantic information but less perceptible to human vision \cite{song2024survey}. This characteristic allows subtle modifications in these regions to go undetected by semantic steganalyzers. As a result, high-frequency region semantic steganography presents a practical strategy. 
For example, we can apply wavelet transforms to secret images to preserve high-frequency semantic features while eliminating low-frequency components.


\textbf{Pre-trained Model-based Semantic Steganography:}
Compared to single-stage end-to-end training, the pre-trained model-based semantic steganography strategy adopts a two-stage separable training approach. The semantic encoder and decoder focus on efficiently compressing and reconstructing semantic information, while the semantic steganography module concentrates on secret semantic feature embedding and extraction. This separable training strategy avoids direct conflicts between the objectives of the two types of tasks, preventing the image steganography JSCC model from falling into local optima. 
For example, Tang et al. \cite{tang2025towards} employed the DeepJSCC architecture as the trained semantic encoder and decoder, then froze their parameters, and further trained the parameters of INN-based signal steganography modules.

\textbf{Coverless Semantic Steganography:}
Cover-based semantic steganography may disrupt statistical properties, leaving traces of rewriting. In contrast, coverless semantic steganography leverages GAI models to directly produce stego semantic features without requiring any cover images \cite{yu2023cross}. For example, we can construct specialized dictionaries and semantic knowledge bases using GAI models. At the transmitter, secret images are mapped onto semantic labels through these dictionaries, which are then utilized by GAI models to generate realistic-looking semantic features to confuse intelligent eavesdroppers. The receiver can identify these semantic labels to retrieve the hidden information. 



\section{Case Study: A Coverless Steganography SemCom Scheme}

\begin{figure*}
\centering
\includegraphics[width=1\textwidth]{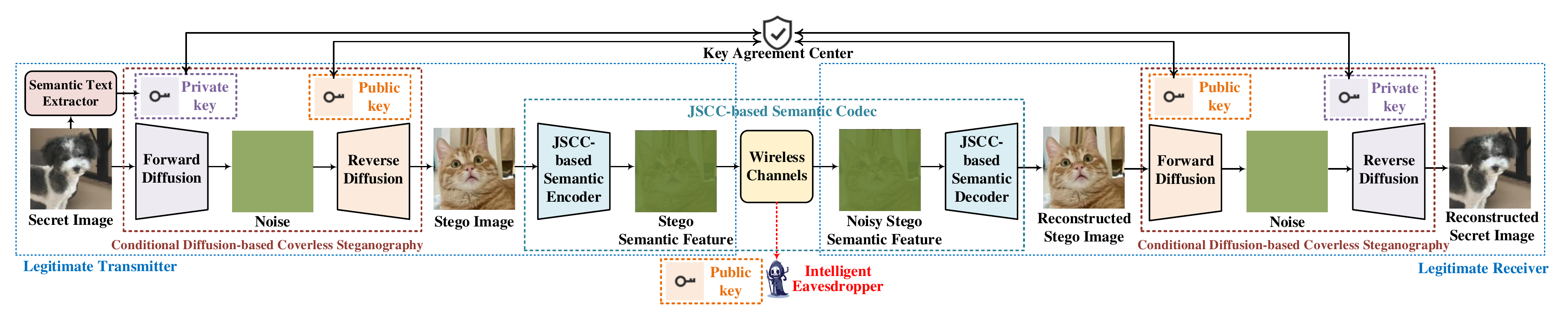}
\caption{Illustration of the proposed coverless steganography SemCom scheme, where the conditional diffusion-based coverless steganography module and JSCC-based semantic codec are trained separately, and keys are generated based on semantic features. The public keys are open access to everyone, including intelligent eavesdroppers, and they are semantic information related to steganographic images but not to secret images.}
\label{figure3}
\end{figure*}

\subsection{Presented Conditional Diffusion-based Coverless Steganography SemCom Scheme}

\begin{figure}
\centering
\includegraphics[width=0.5\textwidth]{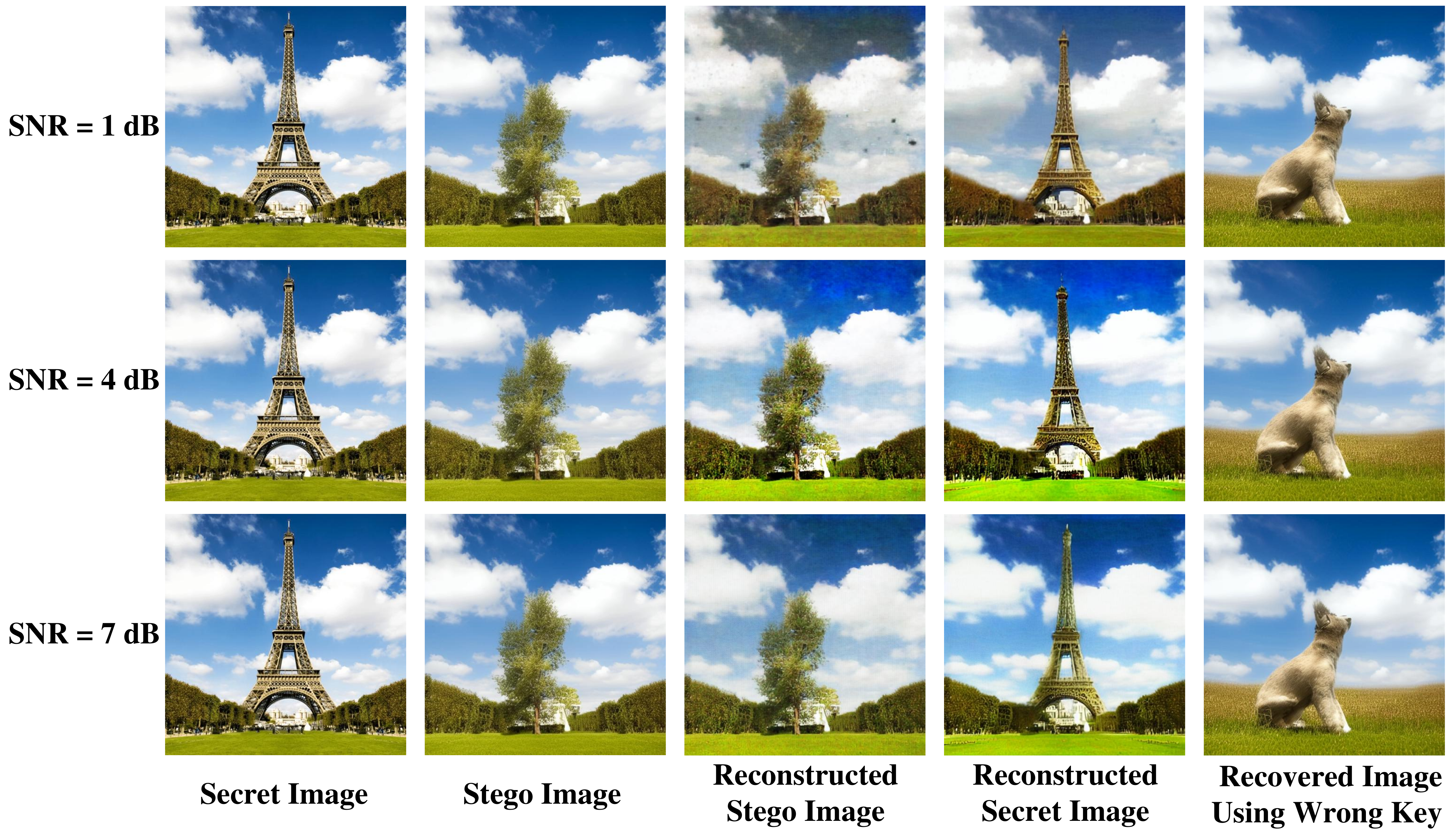}
\caption{Simulation results of the proposed conditional diffusion-based coverless steganography SemCom scheme.}
\label{figure4}
\end{figure}

Existing image steganography SemCom schemes \cite{tang2024secure,tang2025towards,huo2025image,li2024multi} have the following limitations. Firstly, they are heavily constrained by the inherent characteristics of cover images. There is a close positive correlation between the capacity of secret images and the size and quality of cover images. This means that to transmit a high-dimensional secret image, it is necessary to simultaneously obtain a cover image with highly accurate matching. This requirement undoubtedly poses new challenges to resource scheduling and management in SemCom systems. Secondly, the selection and update mechanisms for cover images need further research. If there are insufficient cover images to meet update requirements, or if the update method appears abnormal, it can easily arouse the vigilance and suspicion of intelligent attackers. Thirdly, the embedding of secret semantic features may trigger statistical anomalies. Once intelligent attackers employ advanced steganalysis techniques, these potential statistical anomalies may become the breakthrough point for them to detect semantic steganography behavior.

Against this background, we present a coverless steganography SemCom scheme, as shown in Figure \ref{figure3}. The transmitter and receiver share private and public keys to generate and recover stego images. The key agreement center stores and distributes these keys through secure channels, with each key pair dynamically updated according to the transmitted content. The intelligent eavesdropper is assumed to possess strong reasoning capabilities and access to semantic decoders and various generative AI models, including large language models and diffusion models.In the worst-case scenario, it can intercept the transmitted semantic features, obtain the public key, and infer the complete steganography and recovery processes. Although it may employ these models to reconstruct the secret image, the correct private key remains unavailable.

Specifically, the transmitter first extracts the semantic content of the secret image to generate a private key and derives a public key by modifying its core semantics. The secret image is then mapped into the latent space of a conditional diffusion model, where deterministic DDIM inversion guided by the private key transforms it into a noise representation. Starting from this representation, conditional denoising guided by the public key generates a stego image that presents the public semantics while concealing the secret content. The stego image is subsequently transmitted through a JSCC-based semantic communication system. Upon reception, the legitimate receiver applies DDIM inversion guided by the public key to recover the corresponding noise representation and then performs conditional denoising guided by the private key to reconstruct the secret image. Without the correct private key, the eavesdropper cannot reproduce the private conditional generation process or recover the secret image.



\subsection{Simulation Results}
In our experiments, the datasets are sourced from two publicly available datasets\footnote{\url{https://github.com/aisegmentcn/matting_human_datasets}}\footnote{\url{https://www.kaggle.com/datasets/iamsouravbanerjee/animal-image-dataset-90-different-animals}} and Google search engines.
Stable Diffusion 1.5\footnote{\url{https://huggingface.co/stable-diffusion-v1-5/stable-diffusion-v1-5}} is adopted as the conditional diffusion model, with deterministic DDIM inversion and sampling performed over 50 steps. NTSCC is employed for semantic encoding and decoding, and the stego images are transmitted over wireless channels with different SNRs.


\begin{table*}[htbp]
\centering
\caption{Performance comparison between the legitimate receiver and eavesdropper under different SNRs, where the eavesdropper uses an empty private key without semantic content.}
\label{tab:receiver_eavesdropper_comparison}
\setlength{\tabcolsep}{6pt}
\renewcommand{\arraystretch}{1.15}
\begin{tabular}{c cc cc cc cc cc}
\toprule
\multirow{2}{*}{SNR (dB)}
& \multicolumn{2}{c}{PSNR (dB)}
& \multicolumn{2}{c}{SSIM}
& \multicolumn{2}{c}{LPIPS}
& \multicolumn{2}{c}{MSE}
& \multicolumn{2}{c}{CLIP Score} \\
\cmidrule(lr){2-3}
\cmidrule(lr){4-5}
\cmidrule(lr){6-7}
\cmidrule(lr){8-9}
\cmidrule(lr){10-11}
& Eve & Receiver
& Eve & Receiver
& Eve & Receiver
& Eve & Receiver
& Eve & Receiver \\
\midrule
0  & 18.76 & 19.44 & 0.6189 & 0.6004 & 0.3310 & 0.3372 & 0.01522 & 0.01322 & 24.83 & 28.36 \\
2  & 18.97 & 20.27 & 0.6439 & 0.6308 & 0.3095 & 0.2953 & 0.01459 & 0.01091 & 24.87 & 28.57 \\
4  & 19.04 & 20.63 & 0.6564 & 0.6477 & 0.3018 & 0.2803 & 0.01441 & 0.01014 & 24.82 & 28.58 \\
6  & 19.08 & 20.80 & 0.6638 & 0.6528 & 0.2989 & 0.2802 & 0.01433 & 0.00978 & 24.75 & 28.27 \\
8  & 19.10 & 20.90 & 0.6687 & 0.6587 & 0.2977 & 0.2784 & 0.01429 & 0.00949 & 24.63 & 28.34 \\
10 & 19.11 & 20.80 & 0.6717 & 0.6556 & 0.2975 & 0.2900 & 0.01428 & 0.00974 & 24.68 & 28.16 \\
\bottomrule
\end{tabular}
\end{table*}


Figure \ref{figure4} provides the simulation results of the proposed scheme. The Eiffel Tower serves as the confidential component requiring protection. Consequently, the prompt ``\textit{Eiffel Tower under the blue sky}``functions as the private key, while ``\textit{a tree under the blue sky}`` acts as the public key. Guided by these keys, the legitimate receiver applying the correct private key successfully reconstructs the secret image, whereas an incorrect key fails to produce the intended result.
Table \ref{tab:receiver_eavesdropper_comparison} presents the PSNR, SSIM, LPIPS, and MSE results of the proposed scheme under different SNRs, demonstrating its effectiveness in secret-image recovery. CLIP score is further introduced to evaluate the semantic similarity of the recovered images, where eavesdropper performs worse than the legitimate receivers, confirming the proposed scheme’s resistance to intelligent eavesdropping.


\subsection{Lessons Learned}
When compared to current image steganography SemCom methods, our proposed coverless steganography SemCom approach utilizes conditional diffusion models to produce stego images. This method eliminates the reliance on cover images and offers greater adaptability. Nevertheless, there are several ways in which this scheme could be enhanced: (1) Is it feasible to utilize the forward diffusion module to manage the noise addition process for generating stego images, thereby dispensing with the need for the reverse diffusion module? (2) In situations where multiple semantic features require protection, can we refine the joint conditional input process to derive the joint private keys?

\section{Future Research Directions}
\begin{figure}
\centering
\includegraphics[width=0.5\textwidth]{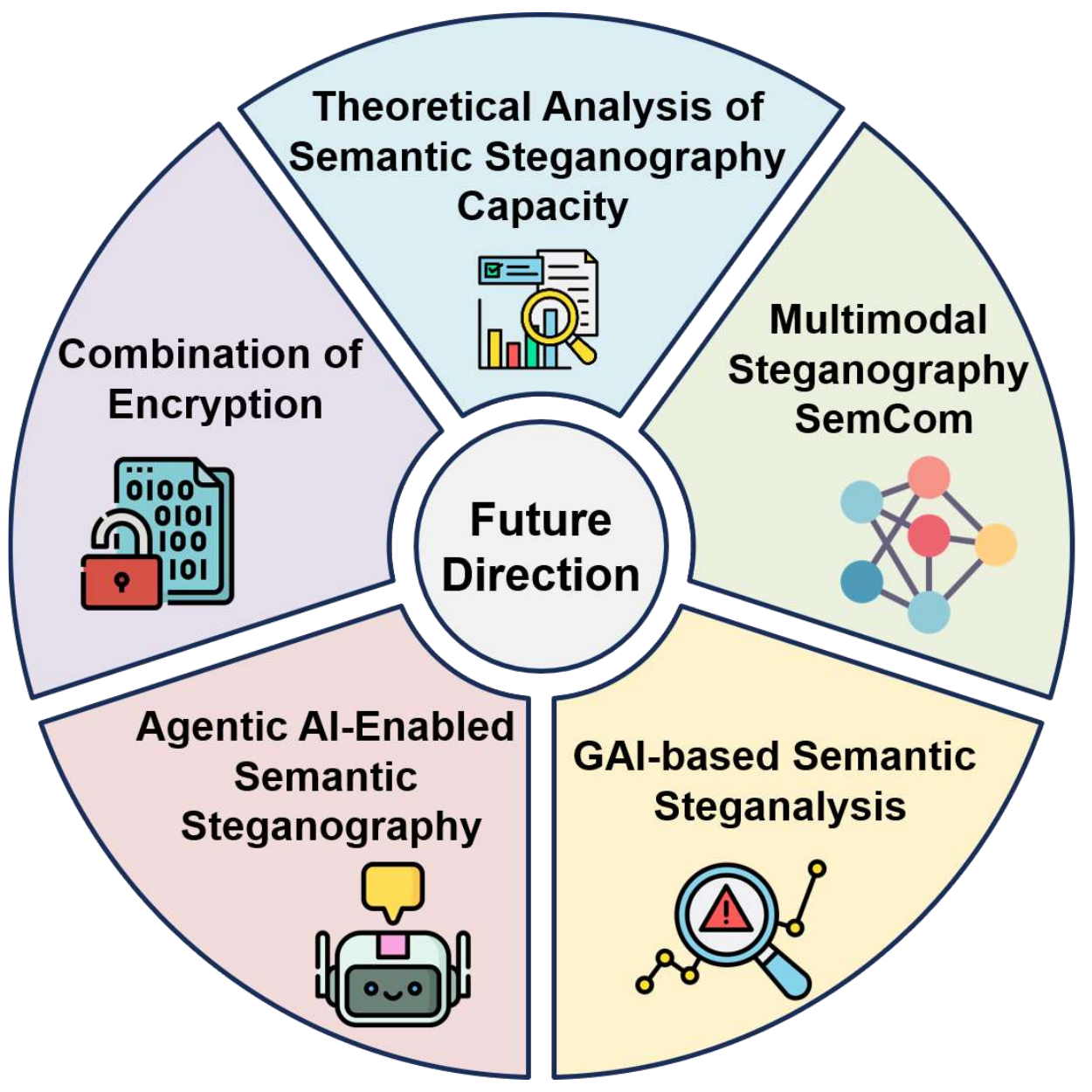}
\caption{Future research directions of semantic steganography communications, which include theoretical analysis of semantic steganography capacity, multimodal steganography SemCom, GAI-based semantic steganalysis, Agentic AI-enabled semantic steganography, and the combination of encryption and steganography} 
\label{future_direction}
\end{figure}

As illustrated in Figure \ref{future_direction}, we discuss five future research directions for securing SemCom, including theoretical analysis of semantic steganography capacity, multimodal steganography SemCom, GAI-based semantic steganalysis, Agentic AI-enabled semantic steganography, and the combination of encryption and steganography.

\subsection{Theoretical Analysis of Semantic Steganography Capacity}
The mathematical theory of SemCom provides the expression of the channel capacity for band-limited Gaussian channels. In steganographic SemCom, when we embed secret semantic features into cover semantic features, the effective channel capacity will decrease compared to SemCom without steganography. This is because, to achieve the steganographic goal and deceive semantic steganalyzers, we need to transmit cover information as well as the correlation information between cover and secret semantic features.
To maximally confuse attackers, reducing the transmission rate of secret semantic features is a direct and effective method. However, doing so will inevitably degrade communication efficiency. Therefore, how to provide a theoretical analysis of semantic steganographic capacity based on channel capacity and steganographic capacity is crucial for guiding further research in the future.

\subsection{Multimodal Steganography SemCom}
Compared to single-image modality SemCom, multimodal SemCom leverages the semantic correlations between multiple modalities to significantly enhance the completeness and precision of semantic understanding. However, the diverse data structures and varying semantic representation methods across different modalities necessitate that steganographic techniques be tailored to the unique characteristics of each modality. Furthermore, given the consistency of semantics across modalities, the stego semantic features must maintain a high degree of uniformity, because any deviation could easily arouse the suspicion of semantic steganalyzers.
To effectively address these challenges, we can employ large multimodal AI models to map data from different modalities into a unified semantic space, within which semantic steganography strategies can be elaborately designed. Concurrently, leveraging the correlations between modalities to embed secret information can further enhance its concealment.

\subsection{GAI-based Semantic Steganalysis}

Traditional discriminative AI achieves steganalysis by learning the classification boundaries between data. However, it is sensitive to local features and prone to missing complex semantic steganography patterns. In contrast, GAI possesses powerful capabilities for modeling complex distributions and can detect traces left by semantic steganography through latent space analysis. From an intelligent attacker's perspective, diffusion models model data distributions through a gradual denoising process, which is expected to capture the latent space shifts caused by semantic steganography. Large language models can also assist attackers in interpreting intercepted content, inferring possible private conditions, and coordinating diffusion models to recover hidden semantic information.
In addition, GAI can automatically synthesize a rich variety of semantic steganography samples or enable adversarial training to effectively improve the generalization performance of detection, making it more advantageous when dealing with various complex semantic steganography scenarios.

\subsection{Agentic AI-Enabled Semantic Steganography}

Existing semantic steganography schemes usually adopt fixed embedding strategies, making it difficult to adapt to changes in communication tasks and channel conditions. Agentic AI can select suitable steganographic methods and adjust embedding parameters according to the required capacity, transmission quality, and security level. It can further assist in understanding task requirements, generating semantic conditions, and coordinating the steganography and communication processes. This enables more flexible semantic steganography with less manual intervention.

\subsection{Combination of Encryption and Steganography for Securing SemCom}

Semantic steganography and semantic steganalysis are always in a state of confrontation, continuously evolving through this dynamic interplay. To provide more reliable privacy protection for SemCom, integrating encryption technology into semantic steganography is a promising solution. Homomorphic encryption allows direct homomorphic operations on encrypted data, ensuring that the privacy of the plaintext is not compromised during processing and providing security protection during semantic extraction and recovery. 
Based on this, homomorphic encryption can be integrated with semantic steganography, enabling semantic extraction and embedding directly on encrypted data without exposing plaintext. Even if attackers extract the hidden ciphertext, they cannot access the secret content without the decryption key.

\section{Conclusions}
In this paper, we have first analyzed the potential of image steganography in securing SemCom. Then, we have provided CNN-based, GAN-based, and INN-based image steganography JSCC models. Furthermore, we have presented six image semantic steganography strategies, including cost-based semantic steganography, adversarial semantic steganography, self learning-based semantic steganography, high-frequency region semantic steganography, pre-trained model-based semantic steganography, and coverless semantic steganography. Additionally, we have given a case study to show the potential of coverless steganography in securing SemCom. Lastly, we have provided future research directions, including theoretical analysis of semantic steganography capacity, multimodal steganography SemCom, GAI-based semantic steganalysis, agentic AI-Enabled Semantic Steganography and combination of encryption and steganography for securing SemCom.

\bibliography{ref.bib}

@article{huo2025image,
  title={Image Semantic Steganography: A Way to Hide Information in Semantic Communication},
  author={Huo, Yanhao and Xiang, Shijun and Luo, Xiangyang and Zhang, Xinpeng},
  journal={IEEE Transactions on Circuits and Systems for Video Technology},
  year={2025},
  volume={35},
  number={2},
  pages={1951-1960},
  publisher={IEEE}
}

@article{zhang2026towards,
  title={Towards Semantic-based Agent Communication Networks: Vision, Technologies, and Challenges},
  author={Zhang, Ping and others},
  journal={SCIENCE CHINA Information Sciences},
  year={2026},
volume={69},
number={10},
pages={200301-},
doi={https://doi.org/10.1007/s11432-026-5054-3}
}

@inproceedings{tang2024secure,
  title={Secure semantic communication for image transmission in the presence of eavesdroppers},
  author={Tang, Shunpu and Liu, Chen and Yang, Qianqian and He, Shibo and Niyato, Dusit},
  booktitle={GLOBECOM 2024-2024 IEEE Global Communications Conference},
  pages={2172--2177},
  year={2024},
  organization={IEEE}
}

@article{tang2025towards,
  title={Towards Secure Semantic Communications in the Presence of Intelligent Eavesdroppers},
  author={Tang, Shunpu and Chen, Yuhao and Yang, Qianqian and Zhang, Ruichen and Niyato, Dusit and Shi, Zhiguo},
  journal={arXiv preprint arXiv:2503.23103},
  year={2025}
}

@article{meng2025survey,
  title={A survey of secure semantic communications},
  author={Meng, Rui and Gao, Song and Fan, Dayu and Gao, Haixiao and Wang, Yining and Xu, Xiaodong and Wang, Bizhu and Lv, Suyu and Zhang, Zhidi and Sun, Mengying and others},
  journal={Journal of Network and Computer Applications},
  pages={104181},
  year={2025},
  publisher={Elsevier}
}

@article{song2024survey,
  title={A survey on Deep-Learning-based image steganography},
  author={Song, Bingbing and Wei, Ping and Wu, Sixing and Lin, Yu and Zhou, Wei},
  journal={Expert Systems with Applications},
  pages={124390},
  year={2024},
  publisher={Elsevier}
}

@article{rustad2023digital,
  title={Digital image steganography survey and investigation (goal, assessment, method, development, and dataset)},
  author={Rustad, Supriadi and Andono, Pulung Nurtantio and Shidik, Guruh Fajar and others},
  journal={Signal processing},
  volume={206},
  pages={108908},
  year={2023},
  publisher={Elsevier}
}

@article{hu2024learning,
  title={Learning-based image steganography and watermarking: A survey},
  author={Hu, Kun and Wang, Mingpei and Ma, Xiaohui and Chen, Jia and Wang, Xiaochao and Wang, Xingjun},
  journal={Expert Systems with Applications},
  pages={123715},
  year={2024},
  publisher={Elsevier}
}

@ARTICLE{guo2024survey,
  author={Guo, Shaolong and Wang, Yuntao and Zhang, Ning and Su, Zhou and Luan, Tom H. and Tian, Zhiyi and Shen, Xuemin},
  journal={IEEE Communications Surveys \& Tutorials}, 
  title={A Survey on Semantic Communication Networks: Architecture, Security, and Privacy}, 
  year={2024},
  doi={10.1109/COMST.2024.3516819}}

@article{yu2023cross,
  title={Cross: Diffusion model makes controllable, robust and secure image steganography},
  author={Yu, Jiwen and Zhang, Xuanyu and Xu, Youmin and Zhang, Jian},
  journal={Advances in Neural Information Processing Systems},
  volume={36},
  pages={80730--80743},
  year={2023}
}

@article{tang2021automatic,
  title={An automatic cost learning framework for image steganography using deep reinforcement learning},
  author={Tang, Weixuan and Li, Bin and Barni, Mauro and Li, Jin and Huang, Jiwu},
  journal={IEEE Transactions on Information Forensics and Security},
  volume={16},
  pages={952--967},
  year={2021},
  publisher={IEEE}
}

@ARTICLE{yang2024secure,
  author={Yang, Zhaohui and Chen, Mingzhe and Li, Gaolei and Yang, Yang and Zhang, Zhaoyang},
  journal={IEEE Network}, 
  title={Secure Semantic Communications: Fundamentals and Challenges}, 
  year={2024},
  volume={38},
  number={6},
  pages={513-520},
  doi={10.1109/MNET.2024.3411027}}

@ARTICLE{shen2024secure,
  author={Shen, Meng and Wang, Jing and Du, Hongyang and Niyato, Dusit and Tang, Xiangyun and Kang, Jiawen and Ding, Yaoling and Zhu, Liehuang},
  journal={IEEE Network}, 
  title={Secure Semantic Communications: Challenges, Approaches, and Opportunities}, 
  year={2024},
  volume={38},
  number={4},
  pages={197-206},
  doi={10.1109/MNET.2023.3327111}}

@inproceedings{li2024multi,
  title={Multi-Modal Task-Oriented Secure Semantic Communication: A Hide-and-Deceive Approach},
  author={Li, Zonglin and Huan, Ouwen and Zhang, Wenjing and Luo, Tao},
  booktitle={2024 10th International Conference on Computer and Communications (ICCC)},
  pages={1477--1482},
  year={2024},
  organization={IEEE}
}

@article{feng2026toward,
  title={Toward intelligent and secure satellite communications: Agentic ai for covert semantics},
  author={Feng, Zhaoxin and Zhang, Ruichen and Qin, Xiaoqi and Xing, Chengwen and Zhao, Nan and Niyato, Dusit and Al-Dhahir, Naofal},
  journal={IEEE Wireless Communications},
  year={2026},
  publisher={IEEE}
}
\bibliographystyle{IEEEtran}

\vfill

\end{document}